\documentclass[12pt]{iopart}

\usepackage{iopams}  
\usepackage{perpage}
\usepackage{color,latexsym,amssymb,amsthm,eufrak,amstext,ifthen}
\begin{document}


\title{A formal introduction to Horndeski and Galileon theories and their generalizations}

\author{C\'edric Deffayet$^1$ and Dani\`ele A.~Steer$^1$}

\address{$^1$ APC (UMR 7164 - APC, Univ Paris Diderot, CNRS/IN2P3, CEA/lrfu, Obs de Paris, 
Sorbonne Paris Cit\'e, France), 10 rue Alice Domon et L\'eonie Duquet, 75205 Paris Cedex 13, France.}

\eads{\mailto{deffayet@iap.fr}, \mailto{steer@apc.univ-paris7.fr}}

\begin{abstract}

We review different constructions of Galileon theories in both flat and curved space, and for both single scalar field models as well as multi-field models.  Our main emphasis is on the formal mathematical properties of these theories and their construction.  

\end{abstract}

\maketitle

\newcommand{\de}{\partial}
\newcommand{\ba}{\begin{eqnarray}}
\newcommand{\ea}{\end{eqnarray}}
\newcommand{\kc}{\kappa_{(5)}}
\newcommand{\kq}{\kappa_{(4)}}
\newcommand{\kcd}{\kappa^2_{(5)}}
\newcommand{\kcq}{\kappa^4_{(5)}}
\newcommand{\kqd}{\kappa^2_{(4)}}
\newcommand{\Lc}{\Lambda_{(5)}}
\newcommand{\Lq}{\lambda_{(4)}}
\newcommand{\MP}{M_{P}}
\newcommand{\MPd}{M_{P}^2}
\newcommand{\Mg}{M_{g}}
\newcommand{\Mf}{M_{f}}
\newcommand{\Mgd}{M_{g}^2}
\newcommand{\Mfd}{M_{f}^2}
\newcommand{\MPmd}{M_{P}^{-2}}
\newcommand{\be}{\begin{equation}}
\newcommand{\ee}{\end{equation}}
\newcommand{\arctanh}{\mathrm{arctanh}}
\newcommand{\BD}{BD}
\newcommand{\AGS}{AGS}
\def\L5{\tilde{\Lambda}}
\def\MM{M_{(4)}}
\def\Wi{W_{\infty}}
\def\Xii{\Xi_{\infty}}
\def\Z{\zeta}
\def\Wo{W_0}
\def\Xio{\Xi_0}

\renewcommand*{\mathellipsis}{%
  \mathinner{{\ldotp}{\ldotp}{\ldotp}}%
}

\newcommand{\CurlyE}{\mathpzc{E}}

\newcommand{\nn}{\nonumber}
\newcommand{\at}{{\mathcal{T}}}
\newcommand{\al}{{\mathcal{L}}}
\newcommand{\Et}{{\mathcal{\cal E}}}
\newcommand{\atold}{{\mathcal{A}}}
\newcommand{\aA}{{\mathcal{A}}}
\newcommand{\ndd}{n}
\newcommand{\npi}{N}
\newcommand{\dani}[1]{\textcolor{red}{\textsc{Dani: #1}}}
\newcommand{\cedric}[1]{\textcolor{blue}{\textsc{Cedric: #1}}}

\section{Flat space-time Galileons in arbitrary dimension}
\subsection{Introduction: the canonical Galileon Lagrangian}
\label{sec:int}

Perhaps the simplest way in which to introduce Galileon models is through a question: for a scalar field $\pi$ in flat space-time, what is the most general theory which has field equations that are polynomial in second order derivatives of $\pi$, do not contain undifferentiated or only once differentiated $\pi$, and do not contain derivatives of order strictly higher than two? The theories which obey these properties are precisely the flat space-time Galileons first presented in the original reference \cite{Nicolis:2008in} and later generalized to curved space-times \cite{Deffayet:2009wt,Deffayet:2009mn}. 

In fact these Galileon theories have a much longer history, having been discovered earlier in different contexts (see Refs.~\cite{Horndeski:1974wa,Fairlie:1992nb,Fairlie:1992he,Fairlie:1991qe}). They are a subset of Horndeski theories \cite{Horndeski:1974wa}, which describe all scalar-tensor theories with second order field equations in curved 4-dimensional space-time. Flat-space time Galileons were also obtained later in a different way by Fairlie {\it et al.}~\cite{Fairlie:1992nb,Fairlie:1992he,Fairlie:1991qe} (see also \cite{Curtright:2012gx}). Here we do not follow these early references, but proceed in the spirit of \cite{Nicolis:2008in,Deffayet:2009wt,Deffayet:2009mn,Deffayet:2011gz}.

In $D$-dimensional flat space-time, Galileons theories can be defined in several ways.  In this section we focus on the simplest Galileon Lagrangian --- the reader is referred to section \ref{sec:other} for other alternative expressions for the flat space-time Galileon Lagrangian.

Consider a Lagrangian of the form 
\ba \label{L0}
\al = \at_{(2n)}^{\mu_{\vphantom{()}1}
\ldots \mu_{\vphantom{()}n}\nu_{\vphantom{()}1}
\ldots \nu_{\vphantom{()}n} } \pi_{\mu_{\vphantom{()}1} \nu_{\vphantom{()}1}} \ldots \pi_{\mu_{\vphantom{()}n} \nu_{\vphantom{()}n}} \ 
\ea
where here, and in the following, successive derivatives of the scalar $\pi$ are denoted by\footnote{Note that, when considered on curved space-time, $\pi_\mu$ will denote the covariant derivative acting on $\pi$, $\nabla_\mu \pi$, and so on for $\pi_{\mu \nu} ...$, i.e. partial derivatives are just to be replaced by covariant derivatives in the notation (\ref{defpimunu}).}
\be \label{defpimunu}
\pi_\mu \equiv \partial_\mu \pi \, , \qquad \pi_{\mu \nu} \equiv \partial_\nu \partial_\mu \pi \, , \qquad
\pi_{\mu \nu \rho } \equiv \partial_\rho \partial_\nu \partial_\mu \pi \, \qquad {\rm etc}.
\ee
The integer $n$ counts the number of twice-differentiated $\pi$'s appearing in the Lagrangian, and the tensor 
$\at_{(2n)}$ appearing in (\ref{L0}) has $2n$ contravariant indices. It is a function of $\pi$ and $\pi_\mu$ only
\ba 
\label{propT}
\at_{(2n)} = \at_{(2n)}(\pi,\pi_\mu) ,
\ea
and is defined to be totally antisymmetric in its first $n$ indices (i.e.~in the $\mu_i$, $1\leq i \leq n$) as well as {\it separately} in its last $n$ indices (i.e.~in the $\nu_i$, $1\leq i \leq n$).
Thus $\al = \al(\pi,\pi_\mu,\pi_{\mu \nu})$, and the corresponding field equations are ${\cal E}=0$ where 
\ba
 {\Et} &\equiv& 
 \left[  \frac{\partial}{\partial \pi} - \partial_\mu \left(\frac{\partial}{\partial \pi_\mu} \right) +  \partial_\mu \partial_\nu \left(\frac{\partial }{\partial \pi_{\mu \nu}} \right) \right] {\cal L}\,
\label{EL}
\ea
 and the term in square brackets is the Euler-Lagrange operator. 
The original Galileon model as defined in Ref.~\cite{Nicolis:2008in} has Lagrangian ${\cal L}^{{\rm Gal},1}$ of the form (\ref{L0}) with 
\ba
\at^{\mu_{\vphantom{()}1}\label{TENSGAL1}
\ldots \mu_{\vphantom{()}n}  \nu_{\vphantom{()}1}
\ldots \nu_{\vphantom{()}n} }_{(2\ndd),\rm{Gal},1}  &\equiv &  {\mathcal A}_{(2\ndd+2)}^{\mu_{\vphantom{()}1}
\ldots \mu_{\vphantom{()}\ndd+1}\nu_{\vphantom{()}1}
\ldots \nu_{\vphantom{()}\ndd+1} } \pi_{\mu_{n+1}} \pi_{\nu_{n+1}} \, ,
\ea
where the $2m$-contravariant tensor ${\mathcal A}_{(2m)}$ is defined by 
\ba \label{DEFAten}
{\mathcal{A}}_{(2m)}^{\mu_{\vphantom{()}1} \mu_{\vphantom{()}2}
\ldots \mu_{\vphantom{()}m} \nu_{\vphantom{()}1} \nu_{\vphantom{()}2}
\ldots \nu_{\vphantom{()}m}} \equiv
\frac{1}{(D-m)!}\,
\varepsilon^{\mu_{\vphantom{()}1}
\mu_{\vphantom{()}2}  \ldots
\mu_{\vphantom{()}m} \sigma_{\vphantom{()}1}\sigma_{\vphantom{()}2}\ldots
\sigma_{\vphantom{()}D-m}}_{\vphantom{\mu_{\vphantom{()}1}}}
\,\varepsilon^{\nu_{\vphantom{()}1} \nu_{\vphantom{()}2} \ldots
\nu_{\vphantom{()}m}}_{\hphantom{\nu_{\vphantom{()}1}
\nu_{\vphantom{()}2} \ldots
\nu_{\vphantom{()}2m}}\sigma_{\vphantom{()}1}
\sigma_{\vphantom{()}2}\ldots \sigma_{\vphantom{()}D-m}} \,.
\ea
Here the totally antisymmetric Levi-Civita tensor is given by
\ba \label{DEFLC}
\varepsilon^{\mu_{\vphantom{()}1} \mu_{\vphantom{()}2} \ldots
\mu_{\vphantom{()}D}} \equiv - \frac{1}{\sqrt{-g}}
\delta^{[\mu_{\vphantom{()}1}}_1 \delta^{\mu_{\vphantom{()}2}}_2
\ldots \delta^{\mu_{\vphantom{()}D}]}_D \, 
\ea
with square brackets denoting unnormalized permutations.
(Notice that definitions (\ref{DEFAten}) and (\ref{DEFLC}) (only) are in fact also valid in arbitrary {\it curved} space-times with metric $g_{\mu \nu}$ and $D\geq m$, and that the metric only enters those definitions via its determinant $g$ and the contraction between $\sigma$ indices.).  
Hence the first form of the Galileon Lagrangian we will consider is given by \cite{Deffayet:2009mn}
\ba \label{LGAL1}
{\cal L}^{\rm{Gal},1}_{\npi} &=& \left({\mathcal A}_{(2\ndd+2)}^{\mu_{\vphantom{()}1}
\ldots \mu_{\vphantom{()}\ndd+1}\nu_{\vphantom{()}1}
\ldots \nu_{\vphantom{()}\ndd+1} }\pi_{\mu_{n+1}} \pi_{\nu_{n+1}} \right) \pi_{\mu_{\vphantom{()}1} \nu_{\vphantom{()}1}} \ldots \pi_{\mu_{\vphantom{()}\ndd} \nu_{\vphantom{()}\ndd}}
\ea
where $N$ indicates the number of times of $\pi$ occurs;
\ba
\npi\equiv \ndd+2  \; (\geq 2),
\nonumber
\ea
and 
\ba
N\leq  D+1
\label{lin}
\ea
 in order for the Lagrangian to be non-zero in $D$-dimensions.  As discussed in Ref.~\cite{Nicolis:2008in}, the Lagrangian is invariant under the
 ``Galilean'' symmetry 
$ \pi_\mu \rightarrow  \pi_\mu + b_{\mu}$,
$\pi \rightarrow \pi + c$ (where
$b_\mu$ and $c$ are constants), 
which is a covariant generalization of the transformation $\pi \rightarrow \pi + b_\mu x^\mu + c$  defined
 in Minkowski spacetime. 

We end this subsection by noting that the Lagrangian ${\cal L}^{\rm{Gal},1}_{\npi}$ can be rewritten in the form
\ba \label{nico1}
{\cal L}^{\rm{Gal},1}_{\npi}&=& - \sum_{\sigma \in S_{\ndd+1}} \epsilon(\sigma)
\bigl[ \pi_{\vphantom{\mu_1}}^{\mu_{\sigma(1)}} \pi_{\mu_1}\bigr]
\bigl[\pi_{\hphantom{\mu_{\sigma(2)}} \mu_2}^{\mu_{\sigma(2)}}
\pi_{\hphantom{\mu_{\sigma(3)}} \mu_3}^{\mu_{\sigma(3)}} \ldots
\pi_{\hphantom{\mu_{\sigma(\ndd+1)}} \mu_{\ndd+1}}^{\mu_{\sigma(\ndd+1)}}\bigr], 
\ea
where $\sigma$ denotes a permutation of signature $\epsilon(\sigma)$ of the permutation group $S_{\ndd+1}$.
This is the original form presented in \cite{Nicolis:2008in}, and the equality of (\ref{LGAL1}) and  (\ref{nico1}) can be shown using the identity 
\ba
\sum_{\sigma \in S_D} \epsilon(\sigma)
g^{\mu_{\sigma(1)}\nu_{\vphantom{()}1}}
g^{\mu_{\sigma(2)}\nu_{\vphantom{()}2}} \ldots
g^{\mu_{\sigma(D)}\nu_{\vphantom{()}D}} =
- \varepsilon^{\mu_{\vphantom{()}1} \mu_{\vphantom{()}2}
\ldots \mu_{\vphantom{()}D}}\,
\varepsilon^{\nu_{\vphantom{()}1} \nu_{\vphantom{()}2}
\ldots \nu_{\vphantom{()}D}}.
\ea

\subsection{Field equations in $D$-dimensions}

At first sight, it is not obvious that the field equations obtained from the Lagrangian (\ref{LGAL1}) (or equivalently (\ref{L0})) are second order --- indeed from (\ref{EL}), one might expect up to forth order derivatives in the equations of motion.  That this is not the case is due to the fact that $\at_{(2n)}$ (or $\aA_{(2n)}$) is totally antisymmetric in its first $n$ indices as well as in its last $n$ indices.  

To see this, initially consider the term $\partial_{\mu} \partial_{\nu} \left( {\partial {\cal L}}/{\partial \pi_{\mu \nu}} \right)$  in the field equations Eq.~(\ref{EL}) arising from the variation of a twice differentiated $\pi$.  For simplicity we work with the form of the Lagrangian given in (\ref{L0}). On focusing on the ``dangerous terms'', namely those containing derivatives of the fields of order 3 or more and indicated by a $\sim$ below, one finds (in the expressions below we only aim to indicate the form of the terms, and are not rigorous with the index structure): 
\ba
\partial_{\mu} \partial_{\nu} \left( \frac{\partial {\cal L}}{\partial \pi_{\mu \nu}} \right) &\sim& \partial_{\mu} \partial_{\nu}  \left(
\at_{(2n)} \pi_{\mu_{\vphantom{()}1} \nu_{\vphantom{()}1}} \ldots \pi_{\mu_{\vphantom{()}k} \nu_{\vphantom{()}k}} \ldots \right)
\nonumber
\\
&\sim& \partial_\mu \left[ \left(\frac{\partial \at_{(2n)}}{\partial \pi_\alpha} \pi_{\alpha \nu} + \frac{\partial \at_{(2n)}}{\partial \pi} \pi_{\nu}\right) \pi_{\mu_{\vphantom{()}1} \nu_{\vphantom{()}1}} \ldots \pi_{\mu_{\vphantom{()}k} \nu_{\vphantom{()}k}} \ldots  
\right.
\nonumber
\\
&&
\qquad + \left. \at_{(2n)} \pi_{\mu_{\vphantom{()}1} \nu_{\vphantom{()}1}} \ldots \pi_{\mu_{\vphantom{()}k} \nu_{\vphantom{()}k} \nu_{\vphantom{()}}} \ldots \right] + {\text{etc}} 
\nonumber
\\
&\sim&
  \frac{\partial \at_{(2n)}}{\partial \pi_\alpha} \pi_{\alpha \nu \mu} \pi_{\mu_{\vphantom{()}1} \nu_{\vphantom{()}1}} \ldots \pi_{\mu_{\vphantom{()}k} \nu_{\vphantom{()}k}} \ldots  
+ \frac{\partial \at_{(2n)}}{\partial \pi_\alpha} \pi_{\alpha \nu} \pi_{\mu_{\vphantom{()}1} \nu_{\vphantom{()}1}} \ldots \pi_{\mu_{\vphantom{()}k} \nu_{\vphantom{()}k} \mu} \ldots  
\nonumber 
\\
&&
+  \frac{\partial \at_{(2n)}}{\partial \pi} \pi_{\nu} \pi_{\mu_{\vphantom{()}1} \nu_{\vphantom{()}1}} \ldots \pi_{\mu_{\vphantom{()}k} \nu_{\vphantom{()}k} \mu} \ldots  
\nonumber 
\\
&& + \at_{(2n)} \ldots \pi_{\mu_{\vphantom{()}l} \nu_{\vphantom{()}l} \mu_{\vphantom{()}}}\pi_{\mu_{\vphantom{()}k} \nu_{\vphantom{()}k} \nu_{\vphantom{()}}}\ldots
+  \at_{(2n)} \ldots \pi_{\mu_{\vphantom{()}k} \nu_{\vphantom{()}k} \nu_{\vphantom{()}} \mu_{\vphantom{()}}} \ldots + {\text{etc}}
\label{hoho}
\ea
 Since derivatives commute on flat-space, and because of the afore-mentioned antisymmetry properties of the indices of $\at_{(2n)}$, the last four terms in (\ref{hoho}) vanish. While the first does not,
it is, however, straightforward to check that an {\it identical} contribution is generated by the second term in the Euler-Lagrange equations (\ref{EL}) (this occurs when the partial derivative $\partial \pi_\mu$ acts on $\at_{(2n)}$): since this appears with the opposite sign, these contributions cancel exactly!   The second term in (\ref{EL}) also generates a term of the form $\pi_{\mu_k \nu_k \mu}$ which, on contraction with $\at_{(2n)}$, vanishes.  

Hence a {\it sufficient} condition for the field equations derived from Lagrangian (\ref{L0}) to stay of order less or equal to $2$ is that the tensor  ${\mathcal T}_{(2\ndd)}^{\mu_{\vphantom{()}1}
\ldots \mu_{\vphantom{()}\ndd}\nu_{\vphantom{()}1}
\ldots \nu_{\vphantom{()}\ndd} }$  is totally antisymmetric in its first $n$ indices as well as separately in its last $n$ indices (see the ``main lemma" of Ref.~\cite{Deffayet:2011gz}). 
Furthermore, one easily concludes from the above and Eq.~(\ref{TENSGAL1}) that
the field equations obtained from ${\cal L}^{\rm{Gal},1}_{\npi}$ read ${\cal E} = -N \times {\cal E}_{N} = 0$ where 
\ba
{\cal E}_{\npi} &=&
-{\mathcal A}_{(2\ndd+2)}^{\mu_{\vphantom{()}1}
\ldots \mu_{\vphantom{()}\ndd+1}\nu_{\vphantom{()}1}
\ldots \nu_{\vphantom{()}\ndd+1} }
\pi_{\mu_1\nu_1} \pi_{\mu_2\nu_2} \ldots \pi_{\nu_{\ndd+1} \mu_{\ndd+1}},
\nonumber
\\
&=& 
\sum_{\sigma \in S_{\ndd+1}} \epsilon(\sigma)
\prod_{i=1}^{\ndd+1} \pi_{\hphantom{\mu_{\sigma(i)}}
\mu_i}^{\mu_{\sigma(i)}}.
\label{eofmG}
\ea
Notice that these equations of motion are of second order only, as advertised, and since they originate from ${\cal L}^{\rm{Gal},1}_{\npi}$ (which contains $N$ factors of $\pi$), they contain $N-1$ factors of $\pi$.

It is interesting to notice from (\ref{lin}) that the largest number of products of fields allowed in $D$-dimensions is $N=D+1$.  In that case, $n+1=D$ so that it follows from the definition of $\aA$ in (\ref{DEFAten}) that ${\cal E}_{D+1}$ is proportional to the determinant of the the matrix of second derivatives $\pi_{\mu \nu}$. Then the equation of motion ${\cal E}_{D+1} = 0 $ is simply the Monge-Amp\`ere equation which has various interesting properties, in particular in relation to integrability (see e.g.~\cite{Fairlie:1994in}).  Also, when $N=D+1$ the Lagrangian
\ba
{\cal L}^{{\rm Gal},1}_{D+1} \propto \det
\left(\begin{array}{cc}\pi_{\mu \nu} & \pi_{\nu} \\ \pi_{\mu} & 0\end{array}\right) \, ,
\nn
\ea
which, when set equal to zero, is the Bateman equation \cite{Fairlie:1994in,Bateman}.

\subsection{Explicit expression for Galileons in $D=4$ dimensions, and some consequences}

When $D=4$ it follows from (\ref{lin}) that $N$, the number of times $\pi$ occurs in the Lagrangian, can take 4 values, $N \in \{2,3,4,5\}$. (Ref.~\cite{Nicolis:2008in} also includes the tadpole $\pi$ in the family of Galileon Lagrangians.) Thus there are only 4 possible non-trivial Galileons Lagrangians of the form (\ref{LGAL1}) in 4-dimensions, and these are given respectively by\footnote{Note that in the expressions (\ref{L2})-(\ref{L5}) there is a global sign difference with respect to the conventions of Ref. \cite{Deffayet:2009wt,Deffayet:2009mn} as well as different global numerical factors (for each ${\cal L}^{{\rm Gal},1}$)  with respect to the convention of Ref. \cite{Nicolis:2008in}.}
\ba \label{LGAL14D}
{\cal L}^{\rm{Gal},1}_{2} &=& {\mathcal A}_{(2)}^{\mu_{\vphantom{()}1}\nu_{\vphantom{()}1} } \pi_{\mu_{1}} \pi_{\nu_{1}}  \nonumber \\
&=& -\pi^\mu \pi_\mu\ \label{L2} \\
{\cal L}^{\rm{Gal},1}_{3} &=& {\mathcal A}_{(4)}^{\mu_{\vphantom{()}1}
 \mu_{\vphantom{()}2}\nu_{\vphantom{()}1}
 \nu_{\vphantom{()}2} } \pi_{\mu_{2}} \pi_{\nu_{2}} \pi_{\mu_{\vphantom{()}1} \nu_{\vphantom{()}1}} \nonumber
\\
&=& \pi^\mu \pi^\nu \pi_{\mu \nu} - \pi^\mu \pi_\mu \Box \pi \label{L3} 
\\
 \nonumber 
{\cal L}^{\rm{Gal},1}_{4} &=& {\mathcal A}_{(6)}^{\mu_{\vphantom{()}1} \mu_{\vphantom{()}2}
 \mu_{\vphantom{()}3}\nu_{\vphantom{()}1}\nu_{\vphantom{()}2}
 \nu_{\vphantom{()}3} } \pi_{\mu_{3}} \pi_{\nu_{3}} \pi_{\mu_{\vphantom{()}1} \nu_{\vphantom{()}1}}  \pi_{\mu_{\vphantom{()}2} \nu_{\vphantom{()}2}}
\nonumber
\\ 
&=&  -\left(\Box \pi\right)^2 \left(\pi_{\mu}\,\pi^{\mu}\right)
+ 2 \left(\Box \pi\right)\left(\pi_{\mu}\,\pi^{\mu\nu}\,\pi_{\nu}\right)
\nonumber 
\\
&& + \left(\pi_{\mu\nu}\,\pi^{\mu\nu}\right) \left(\pi_{\rho}\,\pi^{\rho}\right)
-2 \left(\pi_{\mu}\pi^{\mu\nu}\,\pi_{\nu\rho}\,\pi^{\rho}\right) \label{L4}\\
{\cal L}^{\rm{Gal},1}_{5} &=& {\mathcal A}_{(8)}^{\mu_{\vphantom{()}1}\mu_{\vphantom{()}2}\mu_{\vphantom{()}3}\mu_{\vphantom{()}4}
\nu_{\vphantom{()}1}\nu_{\vphantom{()}2}\nu_{\vphantom{()}3}\nu_{\vphantom{()}4}}\pi_{\mu_{4}} \pi_{\nu_{4}} \pi_{\mu_{\vphantom{()}1} \nu_{\vphantom{()}1}} \pi_{\mu_{\vphantom{()}2} \nu_{\vphantom{()}2}} \pi_{\mu_{\vphantom{()}3} \nu_{\vphantom{()}3}} 
\nonumber
\\
&=& -\left(\Box \pi\right)^3 \left(\pi_{\mu}\,\pi^{\mu}\right)
+ 3 \left(\Box \pi\right)^2\left(\pi_{\mu}\,\pi^{\mu\nu}\,\pi_{\nu}\right)
+ 3 \left(\Box \pi\right) \left(\pi_{\mu\nu}\,\pi^{\mu\nu}\right) \left(\pi_{\rho}\,\pi^{\rho}\right)
\nonumber \\
&& -6 \left(\Box \pi\right)\left(\pi_{\mu}\pi^{\mu\nu}\,\pi_{\nu\rho}\,\pi^{\rho}\right)
-2 \left(\pi_{\mu}^{\hphantom{\mu}\nu}\,\pi_{\nu}^{\hphantom{\nu}\rho}\,\pi_{\rho}^{\hphantom{\rho}\mu}\right) \left(\pi_{\lambda}\,\pi^{\lambda}\right) \nonumber \\
&& -3 \left(\pi_{\mu\nu}\,\pi^{\mu\nu}\right)
\left(\pi_{\rho}\,\pi^{\rho\lambda}\,\pi_{\lambda}\right)
+6 \left(\pi_{\mu}\,\pi^{\mu\nu}\,\pi_{\nu\rho}\,\pi^{\rho\lambda}\,\pi_{\lambda}\right). \label{L5}
\ea
These Lagrangians (\ref{L2})-(\ref{L5}) lead respectively to the field equations ${\cal E}_N=0$ given by Eq.~(\ref{eofmG}), and which read
\ba
{\cal E}_{2} &=& \Box \pi \label{E2} \\
 {\cal E}_{3} &=& \left(\Box \pi\right)^2 -  \pi_{\mu \nu} \pi^{\mu \nu}\label{E3} \\
 {\cal E}_{4} &=& \left(\Box \pi\right)^3 -  3 \Box \pi \pi_{\mu \nu} \pi^{\mu \nu}+ 2 \pi^{\mu}_{\hphantom{\mu} \nu} \pi^{\nu}_{\hphantom{\nu} \rho} \pi^{\rho}_{\hphantom{\rho} \mu}\label{E4}\\
 {\cal E}_{5} &=& \left(\Box \pi\right)^4 -  6 \left(\Box \pi\right)^2 \pi_{\mu \nu} \pi^{\mu \nu}+ 3 \left(\pi_{\mu \nu} \pi^{\mu \nu}\right)^2 \nonumber \\
 && + 8 \left(\Box \pi\right) \pi^{\mu}_{\hphantom{\mu} \nu} \pi^{\nu}_{\hphantom{\nu} \rho} \pi^{\rho}_{\hphantom{\rho} \mu} - 6 \pi^{\mu}_{\hphantom{\mu} \nu} \pi^{\nu}_{\hphantom{\nu} \rho} \pi^{\rho}_{\hphantom{\rho} \sigma} \pi^{\sigma}_{\hphantom{\sigma} \mu}\, . \label{E5}
 \ea

It is interesting to note that the combination of terms appearing in the equations of motion (\ref{E2})-(\ref{E5}) are in fact directly related to the elementary symmetric polynomials of the eigenvalue of the matrix $\pi^\mu_{\;\;\nu}$ see e.g. \cite{Hassan:2011hr}. (Notice that these also appear in the decoupling limit of massive gravity  \cite{deRham:2010kj,deRham:2010ik}.)
 In general, for an arbitrary  $n \times n$ matrix $M^a_{\hphantom{a} b}$ (with $a/b$ a line/column index belonging to $\{1,...,n\}$), the symmetric polynomials $e_k$ with $k=1,2,\ldots,n$ are defined by
\ba
e_k(M) = - \frac{1}{k!}{\mathcal{A}}_{(2k)}^{a_1\cdots a_k b_1\cdots b_k} M_{a_1 b_1} M_{a_2 b_2} \cdots M_{a_k b_k}
\label{NREC}
\ea
so that, in particular, 
\ba 
e_1 \left(M\right) &=& [M] \\
e_2 \left(M\right) &=& \frac{1}{2} \left([M]^2 - [M^2]\right)\\
e_3 \left(M\right) &=& \frac{1}{6} \left([M]^3 - 3 [M][M^2] + 2 [M^3]\right) \\
e_4 \left(M\right) &=& \frac{1}{24} \left([M]^4 - 6[M]^2[M^2] +3[M^2]^2 + 8 [M][M^3]- 6 [M^4]\right)
\ea
where $[M]=M^a_{\hphantom{a} a}$ denotes the trace of $M$.  For an $n \times n$ matrix 
\ba
{\rm det}  (M) = e_n(M) .
\ea
On comparing (\ref{NREC}) with (\ref{eofmG}),
it follows that the equations of motion (\ref{E2})-(\ref{E5}) can be simply rewritten as
\ba
{\cal E}_{k+1} = k! e_k\left(\pi^\mu_{\hphantom{\mu} \nu}\right).
\ea

\subsection{Other form of Lagrangians}
\label{sec:other}

In order to make contact with different formulations of Galileon theories that can be found in the literature, it is useful to note that the Galileon Lagrangian (\ref{LGAL1}) can be written in different, equivalent, ways all of which differ from (\ref{LGAL1}) by an integration by parts.  In this section we again work in an arbitrary number of dimensions $D$.
 
A first possible alternative Lagrangian for the Galileon with, again, $\npi = n+2$ fields is given by
\ba \label{LGAL2}
{\cal L}^{\rm{Gal},2}_{\npi} &=&\left( {\mathcal A}_{(2\ndd)}^{\mu_{\vphantom{()}1}
\ldots \mu_{\vphantom{()}\ndd}\nu_{\vphantom{()}1}
\ldots \nu_{\vphantom{()}\ndd} } \pi_{\mu_1} \pi_{\lambda} \pi^{\lambda}_{\hphantom{\lambda} \nu_1} \right) \pi_{\mu_{\vphantom{()}2} \nu_{\vphantom{()}2}} \ldots \pi_{\mu_{\vphantom{()}\ndd} \nu_{\vphantom{()}\ndd}},\\
&\equiv & \at^{\mu_{\vphantom{()}1}
\ldots \mu_{\vphantom{()}\ndd}  \nu_{\vphantom{()}1}
\ldots \nu_{\vphantom{()}\ndd} }_{(2\ndd),\rm{Gal},2} \pi_{\mu_{\vphantom{()}1} \nu_{\vphantom{()}1}} \ldots \pi_{\mu_{\vphantom{()}\ndd} \nu_{\vphantom{()}\ndd}},
\label{EXEQ21}
\ea
where 
\ba
\at^{\mu_{\vphantom{()}1}
\ldots \mu_{\vphantom{()}\ndd}  \nu_{\vphantom{()}1}
\ldots \nu_{\vphantom{()}\ndd} }_{(2\ndd),\rm{Gal},2} = \frac{1}{n}
\,
{\mathcal A}_{(2\ndd)}^{\alpha_{\vphantom{()}1}
\ldots \alpha_{\vphantom{()}\ndd}\nu_{\vphantom{()}1}
\ldots \nu_{\vphantom{()}\ndd} }
&& \Big[\left( \pi^{\mu_1} \pi_{\alpha_1}\right) \delta^{\mu_2}_{\; \; \alpha_2} \ldots \delta^{\mu_n}_{\; \; \alpha_n}  \nonumber \\
&& + \delta^{\mu_1}_{\; \;  \alpha_1} \left( \pi^{\mu_2} \pi_{\alpha_2} \right) \delta^{\mu_3}_{\; \; \alpha_3} \ldots \delta^{\mu_n}_{\; \;  \alpha_n} \nonumber \\
&& + \ldots \nonumber \\
&& + \delta^{\mu_1}_{\; \;  \alpha_1} \ldots \delta^{\mu_{n-1}}_{\; \;  \alpha_{n-1}} \left( \pi^{\mu_n} \pi_{\alpha_n} \right) \Big].
\label{TT2}
\label{TGAL2}
\ea
Similarly, another integration by parts yields
\ba \label{LGAL3}
{\cal L}^{{\rm Gal},3}_{\npi} &= & \left( {\mathcal A}_{(2\ndd)}^{\mu_{\vphantom{()}1}
\ldots \mu_{\vphantom{()}\ndd}\nu_{\vphantom{()}1}
\ldots \nu_{\vphantom{()}\ndd} } \pi_{\lambda} \pi^{\lambda}\right) \pi_{\mu_{\vphantom{()}1} \nu_{\vphantom{()}1}} \ldots \pi_{\mu_{\vphantom{()}\ndd} \nu_{\vphantom{()}\ndd}}
\ea
so that
\ba
\at^{\mu_{\vphantom{()}1}
\ldots \mu_{\vphantom{()}\ndd}  \nu_{\vphantom{()}1}
\ldots \nu_{\vphantom{()}\ndd} }_{(2\ndd),{\rm Gal},3}  &=& X {\mathcal{A}}_{(2\ndd)}^{\mu_{\vphantom{()}1}
\ldots \mu_{\vphantom{()}\ndd}   \nu_{\vphantom{()}1}
\ldots \nu_{\vphantom{()}\ndd}  }
\ea
where
\ba
X\equiv \pi_\mu \pi^\mu \, .
\label{Xdef}
\ea

The three Lagrangians (\ref{LGAL1}), (\ref{LGAL2}) and (\ref{LGAL3}) are all equal up to a total derivative: from the properties of $\aA_{(2n)}$, it follows that (see \cite{Deffayet:2011gz} for an explicit proof) 
\ba
{\mathcal{L}}_{\npi}^{\rm{Gal},1}&=&\frac{\npi}{2}{\mathcal{L}}_{\npi}^{\rm{Gal},3}-\frac{\npi-2}{2} \partial_\mu J^\mu_{\npi}, \label{REL1}\\
{\mathcal{L}}_{\npi}^{\rm{Gal},1}&=&-\npi {\mathcal{L}}_{\npi}^{\rm{Gal},2}+\partial_\mu J^\mu_{\npi} \, , 
\label{REL3}
\ea
where
the current $J^\mu_{\npi}$ is defined by
\ba \label{DEFJ}
J^\mu_{\npi}= X {{\mathcal{A}}}_{(2n)}^{\mu \mu_{2}\cdots\mu_{n}\nu_{1}\nu_{2}\cdots\nu_{n}}\pi_{\nu_{1}}\pi_{\mu_{2}\nu_{2}}\cdots\pi_{\mu_{n}\nu_{n}} \, .
\ea
Consequently the
 equations of motion of all three Galileon Lagrangians are identical, given by (\ref{eofmG}), and strictly  of second order. 
Finally, observe from (\ref{eofmG}) and (\ref{LGAL3}) that ${\cal L}^{\rm{Gal},3}_{\npi}$ can be rewritten as
\ba
{\cal L}^{\rm{Gal},3}_{\npi} = - X {\cal E}_{\npi-1} 
\label{DEFGAL3}
\ea
where ${\cal E}_{\npi-1}$ are the equations of motion coming from ${\cal L}^{\rm{Gal}}_{\npi-1}$ (where we drop the index $1,2,3$). In this form, one sees directly that Galileon models containing a given number $\npi$ of $\pi$ fields can be obtained from the field equations of the same models with one less field.  It is precisely this property which was used by Fairlie {\it et al.} hierarchical construction of Galileons \cite{Fairlie:1992nb,Fairlie:1992he,Fairlie:1991qe,Curtright:2012gx}.

\section{Generalizing flat space-time Galileons}

As we have explained, by definition, Galileon theories describing a single scalar field $\pi$ have equations of motion which are strictly of order 2 on flat space-time. In this section we focus on three extensions to this scenario.  First, for reasons explained below, we present the most general Lagrangian for $\pi$ which yields equations of motion of order 2 {\it or less}.  Then we discuss multi-galileon scalar theories, as well as $p$-form Galileon theories.

\subsection{Generalized single-field Galileons}
\label{singfield}

The property that flat space-time Galileons have field equations containing second order derivatives only is lost in curved space-time \cite{Deffayet:2009wt} (see also below).  However, it is well known that {\it increasing} the order of the field equations lead to an increase in the number of propagating degrees of freedom.  Hence, when trying to generalize Galileons, it is natural to try to determine the most general scalar theory with field equations containing derivatives of order two {\it or less} on flat space-time. This was done in Ref.~\cite{Deffayet:2011gz} where it was shown that the most  general theory in $D$ space-time dimensions satisfying the three conditions
\begin{itemize}
\item[(i)] its Lagrangian contains derivatives of order 2 or less of the scalar field $\pi$;
\item[(ii)] its Lagrangian is polynomial in the second derivatives of $\pi$;
\item[(iii)] the corresponding field equations are of order 2 or lower in derivatives
\end{itemize}
is given by
\ba
{\cal L} = \sum_{n=0}^{D-1} \tilde{{\cal L}}_{n}\{ f_n \} \, .
\label{final}
\ea
Here the $f_n$ are arbitrary functions of $\pi$ and $X$ (notice that there are $D$ of them);
\ba
\tilde{{\cal L}}_{\ndd}\{f\}&\equiv& f(\pi,X) \, {\cal L}^{\rm{Gal},3}_{\npi = \ndd + 2}, \nonumber \\
&=& f(\pi,X)  \left( X {\mathcal A}_{(2 \ndd)}^{\mu_{\vphantom{()}1}
\ldots \mu_{\vphantom{()}\ndd}\nu_{\vphantom{()}1}
\ldots \nu_{\vphantom{()}\ndd} }  \pi_{\mu_{\vphantom{()}1} \nu_{\vphantom{()}1}} \ldots \pi_{\mu_{\vphantom{()}\ndd} \nu_{\vphantom{()}\ndd}}\right), \label{GENL}
\ea
and the braces indicate that $\tilde{{\cal L}}_{\ndd}\{f\}$ is a functional of $f$. The equations of motion corresponding to each  $\tilde{{\cal L}}_{\ndd}\{f\}$ are\footnote{We use the notation $f_{X} \equiv f_{,X}$, $f_{\pi} \equiv f_{,\pi}$ and so on.}
\begin{eqnarray}
0 & = & 2\left(f+Xf_{X}\right)\mathcal{E}_{N}+4\left(2f_{X}+Xf_{XX}\right)\mathcal{L}_{N+1}^{\text{Gal},2}\nonumber \\
&  & +X\left[2Xf_{X\pi}-\left(n-1\right)f_{\pi}\right]\mathcal{E}_{N-1} \nonumber \\
&& -n\left(4Xf_{X\pi}+4f_{\pi}\right)\mathcal{L}_{N}^{\text{Gal},2} - nXf_{\pi\pi} \mathcal{L}_{N-1}^{{\rm Gal},1}.\label{EOFM}
\end{eqnarray}
where $N=n+2$.   When the function $f$ is constant, these equations reduce to $\mathcal{E}_{N}=0$ as in (\ref{eofmG}). For non-constant $f$, they depend on $\pi^{\mu \nu}$ as well as $\pi^\mu$ through $f$ (which also induces a dependence on $\pi$), $X$ and the different Galileon Lagrangians which appear in (\ref{EOFM}).  Clearly, therefore, the Lagrangian (\ref{final}) is no longer invariant under the ``Galilean" symmetry discussed in section \ref{sec:int}.

\subsection{Multifield theories}
\label{sec3}

Another, different, way in which to generalize flat-space time Galileons is to consider theories with several fields (rather than a single scalar $\pi$) but imposing that the equations of motion are strictly second order. 

One construction of such theories (having some degree of generality) consists in considering sets of $p$-forms $A^a_p$, where $a$ denotes the type of species. Such forms have field strength $F^a_{p+1} = d A^a_p$, such that the exterior derivative $d F^a$ vanishes. Motivated by analysis of section \ref{sec:int} for a single field $\pi$, one can consider actions given by the formal expression  \cite{Deffayet:2010zh} 
\begin{equation}
{\cal L} = \varepsilon^{\mu_1 \mu_2 \dots}
\varepsilon^{\nu_1 \nu_2 \dots\dots}\,
F^a_{\mu_1\mu_2\dots} F^b_{\nu_1 \nu_2 \dots}
\left(\partial_{\mu_k} F^c_{\nu_l \nu_{l+1} \dots}\,\dots
\right)
\left(\partial_{\nu_j} F^d_{\mu_m \mu_{m+1}\dots}\,\dots\right) \, ,
\label{eq7}
\end{equation}
where the different species are labelled by $(a, b, \dots)$.
The number of indices contracted with the first and second
Levi-Civita tensors $\varepsilon$ must be the same and not greater than $D$, but the two terms in brackets
may now involve {\it different} species and therefore a
 {\it different} number of terms.  The Bianchi identities (i.e.~$[d,d]=0$) ensure that only $\partial F$
appears in the field equations, which therefore remain of order two in derivatives.

Various examples of this kind are given in Ref.~\cite{Deffayet:2010zh}. Consider for instance only zero-forms,  say two scalar fields $\pi$ and $\varphi$ with ``field strengths" given respectively by 
$\pi_\mu \equiv \partial_\mu \pi$ and $\varphi_\mu \equiv \partial_\mu \varphi$. On using (\ref{eq7}) one obtains the following Lagrangian
\ba 
{\cal L}(\pi,\varphi) &=& {\mathcal A}_{(2\ndd+2)}^{\mu_{\vphantom{()}1}
\ldots \mu_{\vphantom{()}\ndd+1}\nu_{\vphantom{()}1}
\ldots \nu_{\vphantom{()}\ndd+1} } \pi_{\mu_{n+1}} \varphi_{\nu_{n+1}} \pi_{\mu_{\vphantom{()}1} \nu_{\vphantom{()}1}} \varphi_{\mu_{\vphantom{()}2} \nu_{\vphantom{()}2}} \pi_{\mu_{\vphantom{()}3} \nu_{\vphantom{()}3}} \ldots  \varphi_{\mu_{\vphantom{()}\ndd-1} \nu_{\vphantom{()}\ndd-1}} \pi_{\mu_{\vphantom{()}\ndd} \nu_{\vphantom{()}\ndd}},
\nonumber
\ea
which has strictly second order field equations. Similar models with several scalar fields, generally known as ``multigalileons",  have been constructed in different ways in the literature, for instance by imposing certain internal symmetries (see e.g.~\cite{Padilla:2010de,Padilla:2010tj}), or by using brane-world constructions relying in particular on the analogy between the above described theories and Lovelock actions (see \cite{Deffayet:2010zh}). These latter constructions can then also be used in the curved space-time extensions of Galileons which we will outline below \cite{deRham:2010eu,Hinterbichler:2010xn,Trodden:2011xh,Goon:2011qf,VanAcoleyen:2011mj,Burrage:2011bt,Padilla:2012dx,Gabadadze:2012tr}. 

In the case of a {\it single} $p$-form, we can simply drop the indices $a$, $b$ etc in (\ref{eq7}).   But then, it is possible to show
that the resulting action always leads to vanishing field equations whenever $p$ is odd!  In fact it is not currently known if the equivalent of a Galileon model exists for just a {\it single} odd $p$-form \cite{Deffayet:2010zh}.
However,  a non-trivial theory for several odd or even (possibly single) $p$-forms can be obtained easily --- once again we follow \cite{Deffayet:2010zh} and quote two simple, mixed $0$ \& $1$-form theories. The first Lagrangian, which is linear in the scalar field and quadratic in the gauge field, is defined for any $D\geq 3$:
\ba
{\cal L} = \varepsilon^{\mu\nu\rho}
\varepsilon^{\alpha\beta\gamma}\,
F_{\mu\nu} F_{\alpha\beta}\,
\partial_\rho\partial_\gamma\pi
= 4 F^{\mu\rho} F^\nu_{\hphantom{\nu}\rho} \pi_{\mu\nu}
-2 F^2\, \Box\pi.
\nonumber
\ea
Both its $\pi$ and $A_\lambda$ field equations are obviously of
pure second order, and given by
\begin{eqnarray}
(F_{\mu\nu,\rho})^2
- 2 (F^{\mu\nu}_{\hphantom{\mu\nu},\nu})^2 &=& 0,
\nonumber
\\
F^{\lambda\mu,\nu} \pi_{\mu\nu}
+ F^{\mu\nu}_{\hphantom{\mu\nu},\nu}
\pi^{\lambda}_{\hphantom{\lambda}\mu}
- F^{\lambda\mu}_{\hphantom{\lambda\mu},\mu}\, \Box\pi
&=& 0.
\nonumber
\end{eqnarray}
Similarly, in $D \geq 4$, the mixed model 
\begin{eqnarray}
{\cal L} &=& \varepsilon^{\mu\nu\rho\sigma}
\varepsilon^{\alpha\beta\gamma\delta}\,
\partial_\mu\pi \partial_\alpha\pi\,
\partial_\nu F_{\beta\gamma}\,
\partial_\delta F_{\rho\sigma}\nonumber\\
&=&- 8 (\pi_{\mu} F^{\rho\mu,\nu}
F_{\rho\sigma}^{\hphantom{\rho\sigma},\sigma} \pi_{\nu} )
+ 4 (\pi^{\mu} F_{\mu\nu,\rho})^2
+ 2 (\pi^{\mu} F_{\nu\rho,\mu})^2
\nonumber\\
&&-4 (\pi_{\mu} F^{\mu\nu}_{\hphantom{\mu\nu},\nu})^2
- 2 (\pi_{\mu})^2 (F_{\nu\rho,\sigma})^2
+ 4 (\pi_{\mu})^2 (F^{\nu\rho}_{\hphantom{\nu\rho},\rho})^2
\nonumber
\end{eqnarray}
 which is quadratic in both the scalar field and gauge field also yields pure second order $\pi$ and $A_\lambda$ field
equations:
\begin{eqnarray}
4 (\pi_{\mu\nu} F^{\rho\mu,\nu}
F_{\rho\sigma}^{\hphantom{\mu\sigma},\sigma})
- 2 (F^{\mu\rho,\sigma} \pi_{\mu\nu}
F^\nu_{\hphantom{\nu}\rho,\sigma})
+ 2 (F^{\mu\rho}_{\hphantom{\mu\rho},\rho} \pi_{\mu\nu}
F^{\nu\sigma}_{\hphantom{\nu \sigma}, \sigma})&&
\nonumber\\
- (F_{\rho\sigma,\mu} \pi^{\mu\nu}
F^{\rho\sigma}_{\hphantom{\rho\sigma},\nu})
+ (\Box\pi) (F_{\mu\nu,\rho})^2
- 2 (\Box\pi) (F^{\mu\nu}_{\hphantom{\mu\nu},\nu})^2 &=& 0,
\nonumber\\
2 (\pi_{\mu\rho} F^{\lambda\mu}_{\hphantom{\lambda\mu},\nu}
\pi^{\nu\rho})
+ 2 (\pi^{\lambda\mu}F_{\mu\nu,\rho} \pi^{\nu\rho})
+ 2 (\pi^{\lambda\rho} \pi_{\rho\mu}
F^{\mu\nu}_{\hphantom{\mu\nu},\nu})
- (\pi_{\mu\nu})^2
(F^{\lambda\rho}_{\hphantom{\lambda\rho},\rho})&&
\nonumber\\
- 2 (\Box\pi) (\pi_{\mu\nu} F^{\lambda\mu,\nu})
- 2 (\Box\pi) (\pi^{\lambda}_{\hphantom{\lambda}\mu}
F^{\mu\nu}_{\hphantom{\mu\nu},\nu})
+ (\Box\pi)^2 (F^{\lambda\mu}_{\hphantom{\lambda\mu},\mu})&=& 0.
\nonumber
\end{eqnarray}
Again, similar theories can be obtained following different routes, see for instance \cite{Padilla:2012dx,Zhou:2011ix}.


These models can also be generalized to non-Abelian gauge bosons $A^a_\mu$ and
their field strengths $F = dA + A\wedge A$ \cite{Deffayet:2010zh}. Indeed, if we denote by ${\cal D}$
the gauge covariant derivative, then the Bianchi identities ${\cal
D}^{\vphantom{a}}_{[\mu}F^a_{\nu\rho]} = 0$ still hold. Hence 
Lagrangians of the form ${\cal L}= \varepsilon^{\mu\nu\dots}
\varepsilon^{\alpha\beta\dots}\, F^a_{\mu\nu} F^b_{\alpha\beta}
({\cal D}_\rho F^c_{\gamma\delta}\, \dots) ({\cal D}_\epsilon
F^d_{\sigma\tau}\, \dots)$ define non-linear extensions of
Yang-Mills theory, while keeping field equations of second (and
lower) order (see also \cite{Zhou:2011ix,Goon:2012mu}). In this case, indeed, 
the invariance of the field equations under constant shifts, $A^a_{\mu} \rightarrow
A^a_{\mu} + c^a_{\mu}$ and $F^a_{\mu\nu} \rightarrow F^a_{\mu\nu}
+ k^a_{[\mu\nu]}$ is lost (just because of the form of the field strength and covariant derivative). 
This feature is also shared by the generic models introduced in section \ref{singfield}, where the original ``Galilean" symmetry is lost, as well as with generalized models also introduced in \cite{Deffayet:2010zh} where undifferentiated $F$ also occur in the action.

\section{Covariant Galileons and Horndeski theories}

So far we have worked in flat space-time. In this section we outline the extension of the results presented above to curved $D$-dimensional space-time with metric $g_{\mu \nu}$.  This process is often referred to as ``covariantization''.

\subsection{Introduction with a simple example in $D=4$ dimensions}

Since the expressions we will deal with quickly become complicated, we begin with a specific example.  
Consider the Lagrangian ${\cal L}^{\rm{Gal},1}_4$ of equation (\ref{L4}) in $D=4$ dimensions.  Written on an arbitrary space time (that is, on replacing all partial derivatives by covariant derivatives), the action becomes
\begin{eqnarray} \label{S4}
S_4=\int d^4 x \sqrt{-g}\, {\cal L}^{\rm{Gal},1}_4
\end{eqnarray}
which, on varying with respect to $\pi$, gives the equations of
motion ${\cal E}_4=0$ (where, as previously, an irrelevant numerical factor has been removed), with 
\begin{eqnarray}
{\cal E}_4&\equiv& -\frac{1}{2}\left(\pi_{\mu}\,\pi^{\mu}\right)\left(\pi_{\nu\hphantom{\nu}\rho}^{\hphantom{\nu}\nu\hphantom{\rho}\rho}
- \pi_{\nu\rho}^{\hphantom{\nu\rho}\nu\rho} \right)
-\frac{1}{2}\, \pi^{\mu}\,\pi^{\nu} \left(2\, \pi_{\mu\rho\nu}^{\hphantom{\mu\rho\nu}\rho} - \pi_{\mu\nu\rho}^{\hphantom{\mu\nu\rho}\rho} - \pi_{\rho\hphantom{\rho}\mu\nu}^{\hphantom{\rho}\rho}\right) \nonumber \\
&&-\frac{5}{2} \left(\Box \pi \right) \pi^{\mu}\left(\pi_{\mu\nu}^{\hphantom{\mu\nu}\nu}
- \pi_{\nu\hphantom{\nu}\mu}^{\hphantom{\nu}\nu}\right)
-3 \, \pi_{\mu}\,\pi^{\mu\nu}\left(\pi_{\rho\hphantom{\rho}\nu}^{\hphantom{\rho}\rho}- \pi_{\nu\rho}^{\hphantom{\nu\rho}\rho}\right)
\nonumber \\
&&- 2\, \pi^{\mu}\, \pi^{\nu\rho} \left( \pi_{\nu\rho\mu} - \pi_{\mu\nu\rho} \right)\nonumber \\
&& \left(\Box \pi \right)^3
+2 \left(\pi_{\mu}^{\hphantom{\mu}\nu}\,\pi_{\nu}^{\hphantom{\nu}\rho}\,\pi_{\rho}^{\hphantom{\rho}\mu}\right)
-3 \left(\Box \pi\right) \left(\pi_{\mu\nu}\, \pi^{\mu\nu} \right).
\label{E4a}
\end{eqnarray}
The first two terms contain fourth-order derivatives, the
following three terms contain third-order derivatives and the
last three terms contain second-order derivatives. Of course the fourth and third-order derivatives disappear on a
flat spacetime. Indeed, commuting the derivatives, ${\cal E}_4$ can be rewritten as
\begin{eqnarray}
{\cal E}_4&=&  \left(\Box \pi \right)^3 +2 \left(\pi_{\mu}^{\hphantom{\mu}\nu}\,\pi_{\nu}^{\hphantom{\nu}\rho}\,\pi_{\rho}^{\hphantom{\rho}\mu}\right)
-3  \left(\Box \pi\right) \left(\pi_{\mu\nu}\pi^{\mu\nu} \right)
 +\frac{1}{4}\left(\pi_{\mu}\,\pi^{\mu} \right)\left(\pi_{\nu} \,R^{;\nu}\right) \nonumber \\
&&-\frac{1}{2}\left(\pi_{\mu}\,\pi_{\nu}\,\pi_{\rho} R^{\mu \nu;\rho}\right) -\frac{5}{2} \left(\Box \pi\right)\left(\pi_{\mu}\,R^{\mu \nu}\,\pi_{\nu}\right)
+2 \left(\pi_{\mu}\,\pi^{\mu\nu}\,R_{\nu\rho}\,\pi^{\rho}\right) \nonumber \\&& + \frac{1}{2} \left(\pi_{\mu}\,\pi^{\mu}\right) \left(\pi_{\nu\rho}\,R^{\nu \rho} \right) +2 \left(\pi_{\mu}\,\pi_{\nu}\,\pi_{\rho\sigma}\,R^{\mu \rho \nu \sigma}\right).
\end{eqnarray}
However, one is left over with derivatives of the Ricci tensor and scalar
and hence with third-order derivative of the metric. 

Similarly the stress 
energy tensor $T^{\mu \nu}_4$ (defined in a usual way as $T^{\mu \nu}_4 \equiv (-g)^{-1/2}
\delta S_4/\delta g_{\mu\nu}$, and given explicitly in \cite{Deffayet:2011gz}) contains third-order
derivatives of $\pi$. In fact, these 3rd order derivatives are still there even if flat spacetime $g_{\mu\nu} =
\eta_{\mu\nu}$ were a solution of Einstein's equations. As a result, one sees that once the metric is dynamical, new degrees of 
freedom will propagate even on a Minkowski background. Finally, the energy momentum tensor satisfies the relation 
\begin{eqnarray}
\nabla_\mu T_{4}^{\mu \nu} =2\,\pi^{\nu}\,{\cal E}_4
\label{conservTmunu}
\end{eqnarray}
meaning, in particular, that (on flat space-time) the third derivatives present in
the expression of $T_{4}^{\mu \nu}$ are killed by the
application of an extra covariant derivative. From the above discussion, it is clear that a naive covariantization
leads to higher order derivatives in the equations of motion.

It turns out that a non-minimal coupling of $\pi$ to the metric can simultaneously remove all higher derivatives from the field equations of $\pi$ as well as from the energy momentum tensor \cite{Deffayet:2009wt}. 
Indeed, adding to $S_4$ the action 
\begin{eqnarray} \label{Snm4}
S^\text{nonmin}_4&\equiv&
- \int d^4 x \sqrt{-g} \left(\pi_{\lambda}\,\pi^{\lambda}\right)
\left(\pi_{\mu}\,G^{\mu \nu}\,\pi_{\nu}\right),
\end{eqnarray}
where $G^{\mu \nu}$ is the Einstein tensor,
we obtain the equations of motion for $\pi$ in the form
${\cal E}'_4 = 0$, where ${\cal E}'_4$ is given by
\begin{eqnarray}
{\cal E}'_4 &=& \left(\Box \pi \right)^3 +2  \left(\pi_{\mu}^{\hphantom{\mu}\nu}\,\pi_{\nu}^{\hphantom{\nu}\rho}\,\pi_{\rho}^{\hphantom{\rho}\mu}\right)
-3 \left(\Box \pi\right) \left(\pi_{\mu\nu}\pi^{\mu\nu} \right)
-\frac{1}{2} \left(\Box \pi\right) \left(\pi_{\mu}\,\pi^{\mu}\right) R\nonumber\\
&&- \left(\pi_{\mu}\,\pi^{\mu\nu}\,\pi_{\nu}\right) R
-2\left(\Box \pi\right) \left(\pi_{\mu} \,R^{\mu \nu}\,\pi_{\nu}\right)
+\left(\pi_{\lambda}\,\pi^{\lambda}\right)
\left(\pi_{\mu\nu}\,R^{\mu \nu}\right)\nonumber \\
&&+4 \left(\pi_{\mu}\,\pi^{\mu\nu}\,R_{\nu\rho}\,\pi^{\rho}\right)
+2 \left(\pi_{\mu}\,\pi_{\nu}\,\pi_{\rho\sigma}\,R^{\mu \rho \nu \sigma}\right).
\label{E4prime}
\end{eqnarray}
This equation does not contain derivatives of order
higher than 2, and it obviously reduces to the original form, Eq.~(\ref{E4}), in flat space-time. However, it involves first-order derivatives of $\pi$ in
curved spacetime meaning that Galileon symmetry is broken.  Note also the
complex mixing of the field degrees of freedom implied by the
presence of second derivatives of both $\pi$ and $g_{\mu\nu}$ in
this equation (this has been dubbed ``kinetic gravity braiding'' \cite{Deffayet:2010qz}, and can have important effects in
a cosmological context). Finally, one can show that the action $S^\text{nonmin}_4$ is the unique one which eliminates higher derivatives from both the $\pi$ field equations as well as $T_{4}^{\mu \nu}$ \cite{Deffayet:2009wt,Horndeski:1974wa}. 

 To summarize, ``covariantization'' proceeds as follows: first start with the flat-space Lagrangian and replace partial derivatives by covariant derivatives; then determine the correct ``counterterm(s)'' which remove all higher order derivatives (greater than or equal to 3) in the equations of motion.

\subsection{Covariant Galileons in $D$ dimensions}
 \label{sss}

The results of the previous sub-section can be generalized from
$D=4$ to arbitrary $D$, and also to the other Galilean Lagrangians (with different $n$), see Eq.~(\ref{LGAL1}).

The relevant counterterm is now
\cite{Deffayet:2009mn}
\ba \label{Lnp}
{\cal L}^{{\rm Gal},1}_{(n+1,p)} = {\mathcal A}_{(2 n)}^{\mu_{\vphantom{()}1}
\ldots \mu_{\vphantom{()}\ndd}\nu_{\vphantom{()}1}
\ldots \nu_{\vphantom{()}\ndd} }  \mathcal{R}_{(p)} X^p
\pi_{\mu_{2p+1}} \pi_{\nu_{2p+1}} 
\mathcal{S}_{(q)},
\ea
where $\mathcal{R}_{(p)}$ and $\mathcal{S}_{(q)}$ are defined by
\ba
{\mathcal{R}}_{(p)}&\equiv &\prod_{i=1}^{p}R_{\mu_{2i-1}\mu_{2i}\nu_{2i-1}\nu_{2i}}, \label{barR}
\\
{\mathcal{S}}_{(q)}&\equiv&\prod_{i=0}^{q-1}\pi_{\mu_{n-i}\nu_{n-i}}, \label{barS}
\ea 
with $q= n-1-2p$, and we use the convention that ${\cal{S}}_{1}=\pi_{\mu_n \nu_n} $ and ${\cal{S}}_{q\leq 0} = 1$.  Notice that when $p=0$, there are $q=n-1$ terms in second derivatives of $\pi$, so that ${\cal
L}^{{\rm Gal},1}_{(n+1,0)}\equiv{\cal
L}^{\rm{Gal},1}_{N=n+1}$ given in (\ref{LGAL1}).  For arbitrary $p$, ${\cal
L}^{{\rm Gal},1}_{(n+1,p)}$ is obtained from ${\cal
L}^{{\rm Gal},1}_{(n+1,0)}$ by 
replacing
$p$ pairs of twice-differentiated $\pi$ by a product of
Riemann tensors multiplied by $X$ (with suitable
indices). 
Then \cite{Deffayet:2009mn} the action
\ba
\label{ActionCurvedBackground}
S = \int d^D x \sqrt{-g} \sum_{p=0}^{\lfloor \frac{n-1}{2} \rfloor}
\mathcal{C}_{(n+1,p)} {\cal L}_{(n+1,p)},
\ea
where
the coefficients
$\mathcal{C}_{(n+1,p)}$ are given by
\ba
\mathcal{C}_{(n+1,p)} = \left(-\frac{1}{8}\right)^p
\frac{(n-1)!}{(n-1-2p)!\,(p!)^2}
\ea
and
 $\lfloor \frac{n-1}{2} \rfloor$ is the integer part of $(n-1)/2$,
remarkably leads to field equations both for $\pi$ and the metric
with no more than second derivatives.

\subsection{Covariantized generalized Galileon and Horndeski theories}

A covariantization similar to the one given above also exists for the generalized Galileon of section \ref{singfield}.  Once again, it is not sufficient to simply take the Lagrangian $\tilde{{\cal L}}_{n}\{f\}$ of Eq.~(\ref{GENL}) and replace  all partial derivatives by covariant derivatives: to cancel ``dangerous'' terms in the equation of motion, one must also add the correct finite series of counterterms.
%

These have been determined in \cite{Deffayet:2011gz}, and for a given $n$ (the number of twice-differentiated $\pi$'s appearing in the Lagrangian), the general  term in this series is
\ba \label{defLnp}
\tilde{\mathcal{L}}_{n,p}\{ f\}=\mathcal{A}_{(2n)}^{\mu_{1}\cdots\mu_{n}\nu_{1}\cdots\nu_{n}}\mathcal{P}_{(p)} {\mathcal{R}}_{(p)} {\mathcal{S}}_{(q\equiv n-2p)}
\ea
where we use the convention $\tilde{\mathcal{L}}_{n,0}\{f\}=\tilde{\mathcal{L}}_{n}\{f\}$, the functions $\mathcal{R}_{(p)}$ and $\mathcal{S}_{(q)}$ were defined in Eqs.~(\ref{barR}) and (\ref{barS}), and\footnote{$X_0$ is an arbitrary constant. Its presence is related to the possibility 
of adding terms (all vanishing in flat space) that avoid higher derivatives.}
\ba
\mathcal{P}_{(p)} \equiv \int_{X_{0}}^{X} dX_1 \int_{X_{0}}^{X_{1}} dX_2 \cdots\int_{X_{0}}^{X_{p-1}} dX_p  \; f_n(\pi,X_p)X_p.
\nn
\ea  
In $D$ dimensions, the suitable combination of $\tilde{\mathcal{L}}_{n,p}\{f\}$ eliminating all higher order derivative is found to be given by \cite{Deffayet:2011gz}
\ba
\tilde{\mathcal{L}}^{\rm {cov}}_{n}\{f\}=\sum_{p=0}^{\lfloor\frac{n}{2}\rfloor}\tilde{\mathcal{C}}_{n,p}\tilde{\mathcal{L}}_{n,p}\{f\},
\label{THEANSWER}
\ea
where the coefficients $\tilde{\mathcal{C}}_{n,p}$ are given by\footnote{The difference between $\mathcal{C}_{(n+1,p)}$ and $\tilde{\mathcal{C}}_{n,p}$ arises from the fact that the covariantisation procedure started initially from ${\cal L}^{{\rm Gal},1}$ in section \ref{sss}, whereas here we have started from ${\cal L}^{{\rm Gal},3}$.} 
\ba 
\label{COEFCNP} \tilde{\mathcal{C}}_{n,p}=\left(-\frac{1}{8}\right)^{p}\frac{n!}{(n-2p)!p!}
= \frac{1}{p!} \mathcal{C}_{(n+1,p)}.
\ea  

To summarize, the full covariantized ``generalized Galileon'' Lagrangian in $D$ dimensions is
\ba
{\cal L} = \sum_{n=0}^{D-1} \tilde{{\cal L}}^{\rm cov}_{n}\{ f_n \} \, 
\label{finalb}
\ea
which depends on the $D$ functions $f_n(\pi,X)$, as well as $\sum_{n=0}^{D-1} \lfloor{n/2} \rfloor$ functions of $\pi$ (the integration ``constants'').  However these latter ``constants'' can be re-absorbed by a redefinition of the functional coefficients $f_n$ meaning that the Lagrangian depends on $D$ arbitrary functions only.  

 We end this subsection by comparing the Lagrangian 
(\ref{finalb}) with that constructed by Horndeski in Ref.\cite{Horndeski:1974wa}.  There the author explicitly constructed the most general scalar-tensor theory in 4 dimensions which has field equations of second (and lower) order both for the scalar field and the metric. Using the notation of the present paper, the Horndeski Lagrangian reads
\ba
{\cal L}_{H} &=& -{\cal A}^{\mu_1 \mu_2 \mu_3\nu_1 \nu_2\nu_3}_{(3)} \left(\kappa_1 R_{\mu_1\mu_2\nu_1\nu_2}\pi_{\mu_3\nu_3} - \frac{4}{3} \kappa_{1,X} \pi_{\mu_1\nu_1} \pi_{\mu_2\nu_2} \pi_{\mu_3\nu_3}\right) \label{H11} \\
&& -{\cal A}^{\mu_1\mu_2\mu_3\nu_1\nu_2\nu_3}_{(3)} \left( \kappa_3 R_{\mu_1\mu_2\nu_1\nu_2}\pi_{\mu_3}\pi_{\nu_3} - 4 \kappa_{3,X} \pi_{\mu_1\nu_1} \pi_{\mu_2\nu_2} \pi_{\mu_3}\pi_{\nu_3}\right) \label{H12}\\
&&-{\cal A}^{\mu_1 \mu_2 \nu_1 \nu_2}_{(2)}\left(F R_{\mu_1\mu_2\nu_1\nu_2} - 4 F_{,X} \pi_{\mu_1\nu_1} \pi_{\mu_2\nu_2}\right) \label{H13} \\
&&-2 \kappa_8 {\cal A}^{\mu_1 \mu_2 \nu_1 \nu_2}_{(2)} \pi_{\mu_1}\pi_{\nu_1} \pi_{\mu_2\nu_2} \label{H14} \\&&-3\left(2F_{,\pi} + X \kappa_8\right) X + \kappa_9,\label{H15}
\ea
where $\kappa_1$, $\kappa_3$, $\kappa_8$, $\kappa_9$ and $F$ are  functions of $\pi$ and $X$ which are related by the constraint
\ba
F_{,X} = \kappa_{1,\pi} -\kappa_3 - 2 X \kappa_{3,X}.
\ea
The relation between Horndeski theories (\ref{H11}-\ref{H15}) and theories (\ref{finalb}) with $D=4$ can be summarized as follows. First, observe that the flat space restriction of Horndeski theories must be a subset of the most general flat space theories presented in section  \ref{singfield}.\footnote{Note that this flat space restriction of Horndeski theories is not obviously the most general second order theory for a scalar field in flat space, since the Horndeski construction relies on a condition on the metric field equations which is inapplicable in flat space, and which could reduce the set of theories obtained.} Secondly, the theories (\ref{finalb}) obtained by covariantizing the theory of section \ref{singfield} must be included in the set of theories discussed by Horndeski. In fact they are exactly equivalent \cite{Deffayet:2011gz,Kobayashi:2011nu}, and to see this it is sufficient to rewrite (\ref{H11}-\ref{H15}) in the form (\ref{finalb}):
\ba
{\cal L}_{H}=\sum_{n=0}^{3}\tilde{{\cal L}}^{\text{cov}}_{n}\{f_{n}\}
\label{US15}
\ea
and identify
\begin{eqnarray*}
Xf_{0}(\pi,X) & = & -\kappa_{9}(\pi,X)-\frac{X}{2}\int dX\left(2\kappa_{8}-4\kappa_{3,\pi}\right)_{,\pi},\\
Xf_{1}(\pi,X) & = & X\left(4\kappa_{3,\pi}+\kappa_{8}\right)-\frac{1}{2}\int dX\left(2\kappa_{8}-4\kappa_{3,\pi}\right)+6F_{,\pi},\\
Xf_{2}(\pi,X) & = & 4\left(F+X\kappa_{3}\right)_{,X},\\
Xf_{3}(\pi,X) & = & \frac{4}{3}\kappa_{1,X}.
\end{eqnarray*}

We would like to stress that the constructions of \cite{Deffayet:2011gz} and  \cite{Horndeski:1974wa} are based on very different starting hypotheses.  Ref.~\cite{Deffayet:2011gz} starts from the most general set of scalar theories in flat $D$-dimensional space-time, and then constructs its (possibly non-unique) covariantized generalisation (resulting in the Lagrangian (\ref{finalb})).  Ref.~\cite{Horndeski:1974wa} on the other hand determines the unique set of curved  space-time scalar-tensor theories in 4 dimensions with second order field equations both for the scalar and the metric.  That these two rather different routes end up with {\it identical} sets of theories when $D=4$ is remarkable!  It should be stressed that whether or not this is still true for arbitrary $D$ is still a subject of research, since the extension of Horndeski's construction to $D>4$ is currently unknown.

To finish, notice also that the parametrization given by Horndeski can be somewhat misleading because one might conclude that e.g. the terms (\ref{H11}), (\ref{H13}) and (\ref{H15}),  which all depend on the function $\kappa_1$, are not independent (the same can be said for (\ref{H12}), (\ref{H13}) and (\ref{H15}) which depend on $\kappa_3$). This is, however, not the case, and one can see easily (see \cite{Deffayet:2009mn,Deffayet:2011gz}) that  each of the terms (\ref{H11}-\ref{H15}) lead {\it separately} to field equations of second order both for the scalar and the metric. This 
is transparent in the  rewriting (\ref{US15}). Also, Horndeski's parametrization does not elucidate the relation between the two 
 families of Lagrangians ${\cal L}^{{\rm Gal},1}$ and ${\cal L}^{{\rm Gal},3}$ which appear respectively in Eqs (\ref{H12}),(\ref{H14}) and Eqs (\ref{H11}),(\ref{H13}),(\ref{H15}).

\subsection{Galileons and generalized galileons from branes}

As we have discussed above, when $D=4$ the covariantised generalised Galileon and Horndeski theories are identical, and their Lagrangian therefore describes all scalar-tensor theories in 4 dimensions with second order field equations both for the scalar and the metric.  It therefore follows that any other constructions of Galileons in $D=4$ curved space must be included in Horndeski theories, as a subcase of the action given in Eqs.~(\ref{H11})-(\ref{H15}) (or equivalently (\ref{finalb})).    

This includes in particular the brane constructions of Refs.~\cite{deRham:2010eu,Hinterbichler:2010xn,Trodden:2011xh,Goon:2011qf,Burrage:2011bt}. However, the advantage of such (less general) approaches is that they can give insight into different properties of the theories, for example highlighting the origins of some of the metric-scalar field couplings which appear in Eqs (\ref{H11})-(\ref{H15}). They can also give a geometric interpretation of the Galileon field $\pi$ (or its multifield extensions) and of its galilean symmetry in flat space-time, as well as allowing for an extension of this symmetry to curved (maximally symmetric) backgrounds.  Here we comment briefly on these brane world constructions, without entering into a detailed description.   

The idea  is to consider a 3+1 dimensional brane, our universe, evolving in a higher dimensional space-time (the bulk with metric $G_{AB}$). In the simplest case this is 5 dimensional (so that $A=0,\ldots,4$), but it can be of higher dimensions --- in particular if one is interested in multi-galileons (see e.g.~\cite{Hinterbichler:2010xn}).  Let $X^A(x)$ with denote the brane embedding, with $x^\mu$ ($\mu=0,1,2,3$) the world-volume coordinates.  Then the induced metric $\gamma_{\mu \nu}(x)$ on the brane, and the extrinsic curvature $K_{\mu \nu}(x)$ are defined in the usual way (see for instance \cite{Carter:2000wv}) as
\ba
\gamma_{\mu \nu}= \frac{\partial X^A}{\partial x^\mu} \frac{\partial X^B}{\partial x^\nu} G_{AB}(X) \, \qquad
K_{\mu \nu}= \frac{\partial X^A}{\partial x^\mu} \frac{\partial X^B}{\partial x^\nu} \nabla_A n_B
\ea
where $n^A$ is the unit normal vector to the brane.  Now work in a gauge in which 
\ba
X^\mu(x)=x^\mu \, , \qquad X^5(x)=\pi(x).
\ea
As a result the bulk is foliated by time-like slices given by the surfaces $X^5(x)$=constant, meaning that $\pi$ can be understood as the transverse position of the brane relative to these time-like slices.

One then considers a 4D probe brane with action consisting of non-trivial Lovelock invariants. These are constructed from the induced metric or 4D boundary terms for bulk Lovelock invariants, and thus contain contributions coming both from the brane induced metric (or ``first fundamental form") and the brane extrinsic curvature (or ``second fundamental form"). The observation of Ref.~\cite{deRham:2010eu} is that these terms are such that the field equations derived from such an action are of second order and fall in the generalized Galileon family. Moreover, whenever the bulk is taken to be simple, e.g.~flat, and the brane chosen such that it gives a simple slicing of the bulk (e.g.~such that the induced metric is just de Sitter), then the bulk isometries become imprinted on the brane world-volume theory in the form of a symmetry analogous to the flat space time Galileon symmetry \cite{Goon:2011qf,Burrage:2011bt}. 
The interesting point though is that $\pi$  has a physical interpretation as the brane position, whilst its non-linear symmetries can be identified from the isometries of the bulk.  This procedure can be generalised to multi-galileon theories by considering branes of higher codimension, e.g.~\cite{Hinterbichler:2010xn}.

\subsection{Covariant multi field theories}
Covariantized versions of the multifield theories introduced in section \ref{sec3}, which maintain the second order nature of the field equations, can also be obtained in a way similar to that introduced for scalar theories: all
possible pairs of gradients, $\partial F^a \partial F^b$,
must be replaced by suitable contractions of the undifferentiated
$F^a F^b$ with the Riemann tensor, and added to the
minimally covariantized flat-space action with suitable
coefficients \cite{Deffayet:2010zh}. One should, however, pay attention to the fact that, in the $p
> 0$ construction,  $\nabla_\mu F_{\alpha\beta\dots}$
are to be distinguished from $\nabla_\alpha F_{\mu\nu\dots}$,
essentially because of their different $\varepsilon$-index
contractions, a distinction irrelevant to the original scalar,
$\pi_{\mu\alpha} = \pi_{\alpha\mu}$, case. One common feature
is that flat-space Galilean invariance is also not restorable by
consistent covariantization (nor should it be expected,  given the absence of
constant vectors or tensors in curved space): the equations now
necessarily depend on both second and first derivatives of the
fields. It is also interesting to note that actions trivial in flat space can have non-trivial,
dynamical, curvature-dependent extensions: consider actions (\ref{eq7}) for any single 
odd $p$ form (i.e.~with just one species of odd p-form), which are vacuous in flat space. Their minimal
covariantizations are both nonvanishing and of third
order. However, one may also add appropriate counterterms
that both remove the offending higher derivatives and remain
non-trivial. Indeed, the simplest case is the lowest Galileon
$D=5$ vector action,
\begin{eqnarray}
S_{4,{\rm vec}}  &=&\int d^5 x\, \varepsilon^{\mu\nu\rho\sigma\tau}
\varepsilon^{\alpha\beta\gamma\delta\epsilon}\,
F_{\mu\nu} F_{\alpha\beta}\, \nabla_\rho F_{\gamma\delta}\,
\nabla_\epsilon F_{\sigma\tau}
\nonumber\\
&=&-\frac{1}{2}\int d^5 x\, \varepsilon^{\mu\nu\rho\sigma\tau}
\varepsilon^{\alpha\beta\gamma\delta\epsilon}\,
F_{\mu\nu} F_{\alpha\beta}\,
F^\lambda_{\hphantom{\lambda}\rho} F_{\delta\gamma}\,
R_{\sigma\tau\lambda\epsilon}.
\label{eq17}
\end{eqnarray}
The last equality in (\ref{eq17}) exhibits the model's
curvature-dependence, and is obtained from the first expression
by parts integration. The third
derivatives in the resulting field equations can be removed by
adding the counterterm
\ba
S_{4,{\rm vec}}^{\rm nonmin}=\int d^5 x\, \varepsilon^{\mu\nu\rho\sigma\tau}
\varepsilon^{\alpha\beta\gamma\delta\epsilon}\,
F_{\mu\nu} F_{\alpha\beta}\,
F^\lambda_{\hphantom{\lambda}\rho} F_{\lambda\gamma}\,
R_{\sigma\tau\delta\epsilon}.
\label{eq18}
\ea
It differs from the action (\ref{eq17}) itself simply by an
overall factor and the index change $\delta \leftrightarrow
\lambda$ in the last two terms. 

\section{Conclusions}
Horndeski theories and their recent rediscovery as Galileons, originating in the DGP model \cite{Dvali:2000hr} and its decoupling limit \cite{Luty:2003vm,Nicolis:2004qq}, have recenty been used in numerous phenomenological applications. Although our aim here is only to cover abstract and formal aspects of these theories, we would like to conclude by briefly mentioning some of these applications.

To start with, one of the virtue of Galileons, as introduced in Ref. \cite{Nicolis:2008in}, is to obtain the equivalent of the DGP self-accelerating phase \cite{Deffayet:2001pu,Deffayet:2000uy} without the presence of a ghost instability. Along a similar line, Galileon and extended Galileon models have been used to produce cosmic acceleration or inflation in a novel way (see e.g. \cite{Kobayashi:2010cm,Deffayet:2010qz,Chow:2009fm,Silva:2009km,Kobayashi:2010wa,Burrage:2010cu}),  get interesting cosmological applications of Null Energy Condition violations \cite{Nicolis:2009qm,Creminelli:2010ba,Creminelli:2012my}, or also investigate the cosmological constant problem \cite{Charmousis:2011bf,Copeland:2012qf}. 
Implications for non-Gaussianities \cite{Mizuno:2010ag,Creminelli:2010qf,DeFelice:2011zh,Kobayashi:2011pc,Libanov:2011bk,RenauxPetel:2011dv,Gao:2011qe,RenauxPetel:2011uk,Gao:2011vs} or bouncing cosmologies have also been studied \cite{Qiu:2011cy,Easson:2011zy}. One of the key feature of these models is of course the presence of derivative self-interactions in the scalar sector which allows also to hide the scalar field, when necessary (see e.g. \cite{Babichev:2011kq} for an application to MOND), in a way analogous to the Vainshtein mechanism of massive gravity \cite{Vainshtein:1972sx,Deffayet:2001uk,Babichev:2010jd,Babichev:2009jt,Babichev:2009us} (see e.g.~\cite{Babichev:2013usa} for  recent review, as well as \cite{Babichev:2009ee,Kaloper:2011qc,Babichev:2011iz,Babichev:2012re,DeFelice:2011th}). Gravitational wave emission has also been studied in the context \cite{deRham:2012fw,Chu:2012kz} of the Vainshtein mechanism. Note, however, that many of these phenomenological applications rely on considering regimes in which non-linear terms in the field equations become large and where one would consequently wish to have a good understanding of the UV completion of these theories. Regarding this point, even though some non renormalization theorem can be evocated \cite{Nicolis:2004qq,Hinterbichler:2010xn} the situation has still to be clarified (see Refs \cite{Antoniadis:2002tr,Kohlprath:2004yw,Kiritsis:2001bc,Adams:2006sv,Dvali:2011uu,Dvali:2011th,Dvali:2010jz,Dvali:2012mx,Dvali:2010ns} some of which deal with related or similar issues concerning the DGP model). To conclude, let us mention that solitonic solutions have also been investigated  \cite{Endlich:2010zj,Padilla:2010ir,Masoumi:2012np}, as well as supersymmetrization \cite{Khoury:2011da,Koehn:2013hk}, and that Galileons and their generalizations can also be used to investigate various issues related to causality and chronology protection \cite{Burrage:2011cr,Evslin:2011rj,Evslin:2011vh}.

\ack

DAS thanks CERN for hospitatility whilst this work was in progress.

\section*{References}

\bibstyle{aps}
\bibliography{bibtexDS}
\end{document}